\documentclass[onepage,12pt,a4paper,pdftex]{article}

\UseRawInputEncoding

\usepackage{graphicx, amsfonts, amsmath, amssymb, amscd,  theorem, exscale,
epic, eepic, epsfig, appendix}

\usepackage{hyperref} %

\usepackage{makeidx}
\makeindex

\newcounter{Figure}
\theoremstyle{plain}
\theoremheaderfont{\scshape}

\newtheorem{Def}{\bf Definition}
\newtheorem{The}{\bf Theorem}

\theorembodyfont{\upshape}

\theoremheaderfont{\scshape\small} \theorembodyfont{\scshape\small}

\newcommand{\real}{ {\mathbb R} }

\newcommand{\slap}{\mbox{$ \triangle \mkern -13mu / \ $}}
\newcommand{\nlap}{\mbox{$ \nabla \mkern -13mu / \ $}}
\newcommand{\dlap}{\mbox{$ div \mkern -13mu / \ $}}
\newcommand{\Dlap}{\mbox{$ D \mkern -13mu / \ $}}

\newcommand{\clap}{\mbox{$ curl \mkern -23mu / \ $}}

\newcommand{\be}{\begin{equation}}
\newcommand{\ee}{\end{equation}}
\newcommand{\bea}{\begin{eqnarray}}
\newcommand{\eea}{\end{eqnarray}}
\newcommand{\beas}{\begin{eqnarray*}}
\newcommand{\eeas}{\end{eqnarray*}}

\newcommand{\cDlap}{\mbox{$ \mathcal{D} \mkern -10mu / \ $}}

\newcommand{\mA}{ {\mathcal A} }

\begin{document}

\begin{center}
{\bf \Large Radiation and Asymptotics for Spacetimes with \\ \vspace{5pt} Non-Isotropic Mass}
\end{center}

\begin{center}
{\bf Lydia Bieri  \footnote{lbieri@umich.edu}} \\ \vspace{0.2cm}
{\itshape Dept. of Mathematics, University of Michigan, Ann Arbor, MI 48109-1120, USA}
\end{center}

\begin{center}
{\itshape Dedicated to Demetrios Christodoulou}
\end{center}

\begin{abstract} 
We derive new results on radiation, angular momentum at future null infinity and peeling for a general class of spacetimes. For asymptotically-flat solutions of the Einstein vacuum equations with a term homogeneous of degree $-1$ in the initial data metric, that is it may include a non-isotropic mass term, we prove new detailed behavior of the radiation field and curvature components at future null infinity. In particular, the limit along the null hypersurface $C_u$ as $t \to \infty$ of the curvature component $\rho = \frac{1}{4} R_{3434}$ multiplied with $r^3$ tends to a function $P(u, \theta, \phi)$ on $\real \times S^2$. When taking the limit $u \to + \infty$ (which corresponds to the limit at spacelike infinity), this function tends to a function $P^+ (\theta, \phi)$ on $S^2$. 
We prove that the latter limit does not have any $l=1$ modes. However, it has all the other modes, $l=0, l \geq 2$. 
Important derivatives of crucial curvature components do not decay in $u$, which is a special feature of these more general spacetimes. We show that peeling of the Weyl curvature components at future null infinity stops at the order $r^{-3}$, that is $r^{-4} |u|^{+1}$, for large data, and at order $r^{- \frac{7}{2}}$ for small data. Despite this fact, we prove that angular momentum at future null infinity is well defined for these spacetimes, due to the good behavior of the $l=1$ modes involved.

\end{abstract}


\tableofcontents

\vspace{1cm}

\section{Introduction}
\label{intro}

In this article, we derive results on the radiation field and asymptotics for asymptotically-flat systems containing a non-isotropic mass term that evolves with retarded time $u$. Whereas we focus on solutions of the Einstein vacuum equations, the main findings will hold as well for systems of the Einstein equations coupled to other matter or energy fields. 
As a major consequence, it follows that for physical systems, peeling of the Weyl curvature components at future null infinity stops at the order $r^{-3}$, that is $r^{-4} |u|^{+1}$. This goes beyond the terms $r^{-4} \log r$ derived by Demetrios Christodoulou in the case treated in \cite{chrIV2000}, and  extends the latter scenario to more general settings. Thus, for these dynamical situations (dependence on $u$) there is a natural ``barrier" for peeling that does not change even if one imposes stronger fall-off conditions on the tail of the initial metric and the second fundamental form. 
Moreover, we show that angular momentum at null infinity is well-defined despite the aforementioned properties.

We consider spacetimes $(M, g)$, which solve the Einstein vacuum (EV) equations 
\be \label{EV}
R_{\mu \nu} \ = \ 0 \ \ 
\ee
with $\mu, \nu = 0,1,2,3$, 
for asymptotically flat initial data $(H_0, \bar{g}_{ij}, k_{ij})$ with $i, j = 1, 2, 3$, where $\bar{g}$ and $k$ are sufficiently smooth and  
for which there exists a coordinate system $(x^1, x^2, x^3)$ in a neighborhood of infinity such that with 
$r = (\sum_{i=1}^{3} (x^i)^2 )^{\frac{1}{2}} \to \infty$, it is, referring to this type of initial data and the corresponding spacetimes as (A):  
\bea
\bar{g}_{ij} \ & = & \ \delta_{ij} \ + \ h_{ij} \  + \ o_3 \ (r^{- \frac{3}{2}})  \label{initialdg1}  \\ 
k_{ij} \ & = & \ o_2 (r^{-\frac{5}{2}})  \label{initialdk1}   
\eea
with $h_{ij}$ being homogeneous of degree $-1$. 
In particular, $h$ may include a non-isotropic mass term $M(\theta, \phi)$ depending on the angles. 
The spacetime metric will include a resulting term, being homogeneous of degree $-1$ with corresponding limit $M(u, \theta, \phi)$ at future null infinity depending on the retarded time $u$.

In their pioneering result \cite{sta}, Demetrios Christodoulou and Sergiu Klainerman proved the global nonlinear stability of Minkowski space using initial data of the following type, referring to this type of initial data and the corresponding spacetimes as (CK): 
\bea
\bar{g}_{ij} \ & = & \ (1 \ + \ \frac{2M}{r}) \ \delta_{ij} \ + \ o_4 \ (r^{- \frac{3}{2}}) \label{safg33} \\ 
k_{ij} \ & = & \  o_3 \ (r^{- \frac{5}{2}}) \ ,  \label{safk33}
\eea
where $M$ denotes the mass and is constant.

The most general spacetimes that have been shown to be stable by the present author in \cite{lydia1, lydia2} have initial data of the following type, 
referring to this type of initial data and the corresponding spacetimes as (B): 
\bea
\bar{g}_{ij} \ & = & \ \delta_{ij} \ + \ 
o_3 \ (r^{- \frac{1}{2}}) \label{LBg}  \\
k_{ij} \ & = & \ o_2 \ (r^{- \frac{3}{2}})   \ .      \label{LBk}  
\eea

{\bf Notation:} By $u$ we denote the optical function introduced in section \ref{set} corresponding to {\itshape minus the retarded time} in Minkowski spacetime, and by $\underline{u}$ the corresponding {\itshape advanced time}. We refer to $u$ just as the {\itshape retarded time} with this sign convention.

\subsubsection{New Results}

{\bf Small data}: Under {\itshape smallness assumptions} on the initial data related to ((\ref{initialdg1}), (\ref{initialdk1})) we prove \cite{lydia5}  estimates for the solution spacetimes (A) via the route of a stability proof that lies between (B) in \cite{lydia1, lydia2} and (CK) in \cite{sta}. Note that the results \cite{lydia1, lydia2} imply the existence of solutions of type (A), but we need to make use of the extra structures in ((\ref{initialdg1}), (\ref{initialdk1})) to prove precise estimates for (A), which is done in \cite{lydia5}. \\ 

{\bf Large data}: For {\itshape large data} and situation (A), the main estimates along null hypersurfaces still hold and we derive precise asymptotics. \\

{\bf Non-zero linear momentum and contribution of $\rho$ to gravitational wave memory}: First, for the initial data we can choose a center-of-mass frame where the linear momentum is zero. Under the smallness assumptions, where the resulting spacetime tends to Minkowski spacetime at infinity, one has trivially zero linear momentum. When we consider large data, we can still start with a center-of-mass frame, and we may consider also a center-of-mass frame in the future, but these frames will typically have a non-zero relative velocity. We find that (A) spacetimes naturally exhibit {\itshape non-zero linear momentum}. In particular, we can state the conservation law for linear momentum which takes into account the 
{\itshape linear momentum radiated away}. We also compute a contribution from corresponding $\rho$ limits at future null infinity $\mathcal{I}^+$ to the {\itshape memory effect}. \\ 

{\bf Main Results} of the present article:  
For (A) spacetimes the following hold: 
\begin{itemize}
\item {\bf Peeling} of the Weyl curvature components at future null infinity {\bf stops} at the order $r^{-3}$, that is $r^{-4} |u|^{+1}$ for large data. For small data, this limit is of the dynamical order $r^{- \frac{7}{2}}$, meaning that the leading order term depends on $u$. These orders are achieved by the curvature component $\beta$ for large, respectively small data. 
\item 
{\bf Dynamical behavior with different properties}: Referring to the notation introduced in the next section \ref{set}, we find {\itshape different behavior and fall-off properties at various levels, in particular, at future null infinity and spacelike infinity}. 
We also {\itshape derive different behavior of crucial curvature components and their derivatives}. These include: 
\begin{itemize}
\item[$\bullet$] The limit $\lim_{C_u, t \to \infty} r^3 \rho = P(u, \theta, \phi)$ tends to a function $P^+ (\theta, \phi)$ on $S^2$ when the retarded time $u \to + \infty$. In (CK) the corresponding limit is a constant (see section \ref{future1}). (Note that we use the opposite sign convention for retarded time. Thus, $u \to + \infty$ corresponds to the limit at spacelike infinity, $u \to - \infty$ to the limit at future timelike infinity.) 
\item[$\bullet$] $\rho - \bar{\rho}$, respectively $P - \bar{P}$, does not decay in retarded time $u$. (Here, $\bar{\rho}$ means the mean value of $\rho$ on $S_{t,u}$, and $\bar{P}$ the mean value of $P$ on $S^2$.) 
\end{itemize}
\item {\bf Limits:} Denote by $P_{H_0} (\theta, \phi)$ the limit of $r^3 \rho$ at spacelike infinity.  {\bf The following limits obey}
\[
P_{H_0} (\theta, \phi) \ \neq \ P^+ (\theta, \phi) \ \ \ \  \mbox{in general} \ , 
\]
however, 
\[
\int_{S^2} P_{H_0} (\theta, \phi) \ = \ \int_{S^2} P^+ (\theta, \phi) \ . 
\] 
\item $\mathbf{P_{H_0} (\theta, \phi)}$, respectively $\mathbf{P^+ (\theta, \phi)}$, do {\bf not have any $l=1$ modes}. 
\item {\bf Energy and momenta} at future null infinity are well-defined. In particular, {\bf angular momentum} can be defined and is finite despite the slow decay for $\beta$ and its derivatives. The reason is that $P^+ (\theta, \phi)$ does not have any $l=1$ modes, and that the $l=1$ modes of the involved quantities behave better than these quantities themselves. 

\end{itemize}

Further, {\bf we find that} (see section \ref{main1} of present article and \cite{lydia5})
\begin{itemize}
\item[$\bullet$] We find that 
$\nlap \rho = O \ ( r^{- 4})$ in (A), whereas in (CK) it is $\nlap \rho = O \ ( r^{- 4} \tau_-^{- \frac{1}{2}})$. 
\item[$\bullet$] $\underline{\beta}_4 = O \ ( r^{- 4})$ in (A) while $\underline{\beta}_4 = O \ ( r^{- 4} \tau_-^{- \frac{1}{2}})$ in (CK), and for (A) we have 
$\beta_3 = O \ ( r^{- 4})$ while for (CK)  $\beta_3 =  O \ ( r^{- 4} \tau_-^{- \frac{1}{2}})$. 
Note that the ``rougher energy estimates" together with the Bianchi equations would only give $\nlap \rho = O \ ( r^{- 3} \tau_-^{- \frac{3}{2}})$, $\underline{\beta}_4 = O \ ( r^{- 3} \tau_-^{- \frac{3}{2}})$, 
$\beta_3 = O \ ( r^{- 3} \ \tau_-^{- \frac{3}{2}})$. A discussion of the energy estimates is given in section \ref{energiescurvature}. 
\end{itemize}

\section{Setting}
\label{set}

Next, we shall introduce the main foliations and the frames that we work with. 
Denote by $t$ a maximal time function foliating the spacetime into spacelike hypersurfaces $H_t$. In this framework, the zero-coordinate will be the time-coordinate, and indices $1, 2, 3$ refer to spatial coordinates. Denote by $u$ the optical function (retarded time with the sign convention above) and by $C_u$ the outgoing null hypersurfaces of the $u$-foliation. We denote the corresponding intersection by $S_{t,u} = H_t \cap C_u$. The $S_{t,u}$ are diffeomorphic to the sphere $S^2$ and we refer to $\theta, \phi$ given on $S_{t,u}$ as the spherical variables. 
In particular, the optical function $u$ solves the eikonal equation 
$g^{\alpha \beta} \frac{\partial u}{\partial x^{\alpha}} \frac{\partial u}{\partial x^{\beta}} = 0$. 
We have 
$l^{\alpha} = - g^{\alpha \beta} \partial_{\beta} u$, and the integral curves of $l$ are null geodesics. 
The null hypersurfaces $C_u$ are generated by null geodesic segments. 
Quantities with an overline refer to $H_t$. The covariant differentiation on the spacetime $M$ is written as $D$ or $\nabla$, and the one on a spacelike hypersurface $H$ is $\overline{\nabla}$ or $\nabla$. It is clear from the context what $\nabla$ refers to. 
Now, we introduce 
$\underline{u} := 2r - u$ 
with $r = r(t,u)$ being defined by $4 \pi r^2$ expressing the surface area of $S_{t,u}$. 
Denote by $\underline{C}_{\underline{u}}$ the incoming null hypersurfaces. 
Define $\tau_- := \sqrt{1 + u^2}$, and $\tau_+ := \sqrt{1 + \underline{u}^2}$.

We also work with the null frame $e_4, e_3, e_2, e_1$, where $e_4$ (outgoing null direction) and $e_3$ (ingoing null direction) form a 
null pair that is supplemented by $e_A, \ A = 1, 2$ being a local 
frame field for $S_{t,u} = H_t \cap C_u $. We have $g(e_4, e_3) = -2$. Given this null pair, $e_3$ and $e_4$, we
can define the tensor 
of projection from the tangent space of $M$ to that of $S_{t,u}$ 
\[  \Pi^{\mu \nu} = g^{\mu \nu} + 
\frac{1}{2}(e_4^{\nu}e_3^{\mu}+e_3^{\nu}e_4^{\mu}). \] 
Denote by $N$ the outward unit normal vector of $S_{t,u}$ in $H_t$ and by $T$ the future-directed unit normal to $H_t$, that is $T = \frac{1}{\Phi}  \frac{\partial}{\partial t}$, where $\Phi$ is the lapse function. 
Then we see that $e_3 = T - N$ is an incoming null vectorfield, and $e_4 = T + N$ an outgoing null vectorfield. 
We make use of the expression 
$N = a^{-1} \frac{\partial}{\partial u}$ 
with lapse $a = |\nabla u|^{-1}$. 
Operators on the surfaces $S_{t,u}$ are written with a slash, thus $\dlap$, $\clap \ \ $ are the corresponding divergence and curl operators, respectively. 
For a $p$-covariant tensor field $t$ that is tangent to $S$ we denote by $\Dlap_4 t$ and $\Dlap_3 t$ the projections to $S$ of 
$D_4 t$, respectively $D_3 t$.

Next, we define the components of the Weyl curvature with respect to the null foliation. 

\begin{Def}
We define the null components of the Weyl curvature $W$ as follows: 
\bea
\underline{\alpha}_{\mu \nu} \ (W) \ & = & \ 
\Pi_{\mu}^{\ \rho} \ \Pi_{\nu}^{\ \sigma} \ W_{\rho \gamma \sigma \delta} \
e_3^{\gamma} \ e_3^{\delta} 
\label{underlinealpha} \\ 
\underline{\beta}_{\mu} \ (W) \ & = & \ 
\frac{1}{2} \ \Pi_{\mu}^{\ \rho} \ W_{\rho \sigma \gamma \delta} \  e_3^{\sigma} \
e_3^{\gamma} \ e_4^{\delta} 
\label{underlinebeta} \\ 
\rho \ (W) \ & = & \ 
\frac{1}{4} \ W_{\alpha \beta \gamma \delta} \ e_3^{\alpha} \ e_4^{\beta} \
e_3^{\gamma} \ e_4^{\delta} 
\label{rho} \\ 
\sigma \ (W) \ & = & \ 
\frac{1}{4} \ \ ^*W_{\alpha \beta \gamma \delta} \ e_3^{\alpha} \ e_4^{\beta} \
e_3^{\gamma} \ e_4^{\delta} 
\label{sigma} \\ 
\beta_{\mu}  \ (W) \ & = & \  
\frac{1}{2} \ \Pi_{\mu}^{\ \rho} \ W_{\rho \sigma \gamma \delta} \ e_4^{\sigma} \
e_3^{\gamma} \ e_4^{\delta} 
\label{beta} \\ 
\alpha_{\mu \nu} \ (W) \ & = & \ 
\Pi_{\mu}^{\ \rho} \ \Pi_{\nu}^{\ \sigma} \ W_{\rho \gamma \sigma \delta} \
e_4^{\gamma} \ e_4^{\delta}  \ . 
\label{alphaR}
\eea
\end{Def}

Thus, capital indices taking the values $1,2$, we have: 
\bea
W_{A3B3} \ & = & \ \underline{\alpha}_{AB} \label{intnullcurvalphaunderline*1} \\ 
W_{A334} \ & = & \ 2 \ \underline{\beta}_A \\ 
W_{3434} \ & = & \ 4 \ \rho \\ 
\ ^* W_{3434} \ & = & \ 4 \ \sigma \\ 
W_{A434} \ & = & \ 2 \ \beta_A \\ 
W_{A4B4} \ & = & \ \alpha_{AB}  \label{intnullcurvalpha*1}
\eea
with \\ 
\begin{tabular}{lll}
$\alpha$, $\underline{\alpha}$ & : & $S$-tangent, symmetric, traceless tensors \\ 
$\beta$, $\underline{\beta}$ & : &  $S$-tangent $1$-forms \\ 
$\rho$, $\sigma$ & : & scalars \ . \\ 
\end{tabular}

The Weyl tensor $W_{\alpha \beta \gamma \delta}$ is decomposed into its electric and magnetic parts, which are defined by
\begin{eqnarray}
{E_{ab}} := {W_{aTbT}} \label{Wel1}
\\
{H_{ab}} := {\textstyle {1 \over 2}} {{\epsilon ^{ef}}_a}{W_{efbT}} \label{Wma1}
\end{eqnarray}
Here $\epsilon _{abc}$ is the spatial volume element and is related to the spacetime volume element by
${\epsilon_{abc}} = {\epsilon_{Tabc}}$.

In particular, in our notation it is 
\beas
E_{NN} \  =  \ \rho \ \ \ \ & , & \ \ \ \ H_{NN} \ = \ \sigma \ . 
\eeas

We use a maximal time function. Thus, 
consider a foliation by a maximal time function (maximal foliation): 

{\itshape Constraint equations for a maximal foliation:}
\bea
tr k \ & = & \ 0  \label{k1}  \\ 
\nabla^i  k_{ij}  \ &  = & \ 0  \label{intmaxequs2} \\ 
\bar{R}  \ & = & \  |k|^2  \ \ . \label{intmaxequs3}
\eea

{\itshape Evolution equations for a maximal foliation:}
\bea
\frac{\partial \bar{g}_{ij}}{\partial t} \ & = & \ - 2 \Phi k_{ij}  \label{intmaxequs4} \\ 
\frac{\partial k_{ij}}{\partial t} \ & = & \  - \nabla_i  \nabla_j \Phi  \label{intmaxequs5}
\ + \ (\bar{R}_{ij} \  - \ 2 k_{im} k^m_j) \ \Phi \ \ .  \label{intmaxequs6}
\eea

In view of the maximality condition, taking the trace of the second variation equations 
(\ref{intmaxequs6}), yields the 
{\itshape lapse equation:}
\bea 
\triangle \ \Phi \ & = & \ \mid k \mid^2 \ \Phi \ \ . \label{intmaxequs7}
\eea

From (\ref{initialdk1}) and (\ref{intmaxequs3}) it follows that 
\be 
\bar{R}  \ =  \  |k|^2 \ = \ o(r^{-5})   \label{intRbar1} \ \ . 
\ee
The metric (\ref{initialdg1}) would only yield $\bar{R} = O(r^{-3})$. 
Thus the $r^{-3}$ part of $ \bar{R}$ has to vanish. 
This gives an equation for $h_{ij}$, namely 
\be \label{barR*1}
\bar{R} \ = \ \frac{1}{2} \big( \partial_i \partial_j h_{ij} - \partial_j \partial_j h_{ii}  \big) 
+ o(\mid x \mid^{- \frac{7}{2}}) \ \ . 
\ee
Thus, the first term on the right hand side of (\ref{barR*1}) must vanish separately: 
\be \label{didjhij*1}
\partial_i \partial_j h_{ij} - \partial_j \partial_j h_{ii}  \ = \ 0  \ \ . 
\ee

Moreover, in our setting, in each spacelike $H_t$ of the spacetime $M$ we have the following equations 
\bea
curl \  k \ & = & \ H   \label{k3}  \\ 
\bar{R}_{ij} \ & = & \ k_{im} k^m_j \ + \ E_{ij}  \ .  \label{k4} 
\eea
The detailed equations involving all the components show interesting structures.

\section{Main Results}
\label{main1}

\subsection{Spacetime Behavior}

\subsubsection{Curvature}
\label{curvature1}

In order to put our new results in context, we recall the following two sets of results. 

From \cite{lydia1, lydia2} it follows for spacetimes of type (B)
\bea
\underline{\alpha} \ & = & \ O \ ( r^{- 1} \ \tau_-^{- \frac{3}{2}}) \\ 
\underline{\beta} \ & = & \ O \ ( r^{- 2} \ \tau_-^{- \frac{1}{2}}) \\ 
\rho , \ \sigma , \ \alpha , \ \beta \ & = & \ o \ (r^{- \frac{5}{2}})  
\eea

From \cite{sta} it follows for spacetimes of type (CK)
\bea
\underline{\alpha} \ & = & \ O \ ( r^{- 1} \ \tau_-^{- \frac{5}{2}}) \\ 
\underline{\beta} \ & = & \ O \ ( r^{- 2} \ \tau_-^{- \frac{3}{2}}) \\ 
\rho 	\ & = & \ O \ ( r^{- 3})  \label{rhoCK1} \\ 
\rho - \bar{\rho} \ & = & \ O \ ( r^{- 3} \ \tau_-^{- \frac{1}{2}}) \label{rhoCK2} \\ 
\sigma \ & = & \ O \ ( r^{- 3} \ \tau_-^{- \frac{1}{2}}) \\ 
\sigma - \bar{\sigma}  \ & = & \ O \ ( r^{- 3} \ \tau_-^{- \frac{1}{2}}) \\ 
\beta \ & = & \ o \ (r^{- \frac{7}{2}}) \label{betaCK} \\ 
\alpha \ & = & \ o \ (r^{- \frac{7}{2}})
\eea

We find interesting dynamics in the new situation that are different from the situation (CK) studied in \cite{sta}. 
Namely, in the {\bf present situation}, it follows that for {\bf spacetimes of type (A)} it is 
\bea
\underline{\alpha} \ & = & \ O \ ( r^{- 1} \ \tau_-^{- \frac{5}{2}}) \\ 
\underline{\beta} \ & = & \ O \ ( r^{- 2} \ \tau_-^{- \frac{3}{2}}) \\ 
\rho 	\ & = & \ O \ ( r^{- 3})  \label{rhoA1} \\ 
\rho - \bar{\rho} \ & = & \ O \ ( r^{- 3})  \label{rhoA2} \\ 
\sigma \ & = & \ O \ ( r^{- 3} \ \tau_-^{- \frac{1}{2}}) \\ 
\sigma - \bar{\sigma}  \ & = & \ O \ ( r^{- 3} \ \tau_-^{- \frac{1}{2}}) \\ 
\beta \ & = & \ o \ (r^{- \frac{7}{2}})   \label{betaA}  \\ 
\alpha  \ & = & \ o \ (r^{- \frac{7}{2}})  
\eea

In particular, we also derive for (A) 
\bea
\nlap \rho 	\ & = & \ O \ ( r^{- 4}) \label{nablarhoLB1}
\eea
whereas in \cite{sta} for (CK) by Christodoulou and Klainerman it is 
\bea
\nlap \rho 	\ & = & \ O \ ( r^{- 4} \tau_-^{- \frac{1}{2}})  \label{nablarhoCK}  \ . 
\eea

In \cite{chrIV2000} Christodoulou considered data with slightly stronger decay than in \cite{sta}, namely the tail of the initial data metric takes an extra $r^{- \frac{1}{2}}$. He then shows that the $\beta$ curvature component takes a logarithmic term, in particular it contains a term of the order $1/(r^4 \log r)$. He then discusses this situation in a physical content. Christodoulou finds that 
\bea
\nlap \rho 	\ & = & \ O \ ( r^{- 4} \tau_-^{- 1})  \label{nablarhoDC1} \ . 
\eea

Note that in the Newman-Penrose picture with stronger decay, $\beta$ would peel like $r^{-4}$ and $\alpha$ like $r^{-5}$. Christodoulou's article \cite{chrIV2000} shows that this does not occur for physical data. The present article and a forthcoming paper by the present author further show that dynamical spacetimes do not necessarily exhibit the full peeling, rather ``spherical symmetry" adds to more decay (in the sense of (\ref{rhoCK2}), compare with (\ref{rhoA2})), whereas the curvature and derivatives in spacetimes for more general data tend to fall off more slowly ((\ref{nablarhoLB1}), (\ref{nablarhoCK})). 

We point out that the differences between (\ref{rhoCK2}) and (\ref{rhoA2}) are crucial. Thus, whereas in (CK) $\rho - \bar{\rho}$ takes extra decay in $\tau_-$, this is not the case for (A). The reason for the latter is that the mass depends on the angles, whereas it is a constant in the former case. 

The asymptotic spherical symmetry of the (CK) spacetimes, that roots in the (CK) data (\ref{safg33}), that is $M$ is a constant, allows in \cite{sta} for energies involving rotational vectorfields to control the decay rates of the curvature components. In particular, the extra decay for 
for $\rho - \bar{\rho}$ as in (\ref{rhoCK2}) and for $\nlap \rho$ as (\ref{nablarhoCK}) are obtained. The lack of this symmetry, more precisely the dependence of the mass term in (\ref{initialdg1}) on the angles yields the slower fall-off of these terms in (A). Note that the Bondi mass aspect function $M(\theta, \phi, u)$ is more general in (A). We may think of the (A) spacetimes as the dynamical situations versus solutions with ``non-dynamical" leading order behavior.

\subsubsection{Ricci Coefficients}
\label{Riccicoefficients1}

By $\widehat{\chi}$, $\underline{\widehat{\chi}}$ we denote the shears, which are defined to be the traceless parts of the second fundamental forms 
with respect to the null vectorfields $L$ and $\underline{L}$ generating the corresponding outgoing, respectively incoming null hypersurfaces. Further, $\zeta$ is the torsion-one-form. 
Let $X, Y$ be arbitrary tangent vectors to $S_{t, u}$ at a point. Then
the second fundamental forms are defined as  
\[
\chi (X, Y) = g(\nabla_X L, Y) \ \, \ \ \underline{\chi} (X, Y) = g(\nabla_X \underline{L}, Y) . 
\]
We write for the trace of these tensors $tr \chi$, respectively $tr \underline{\chi}$. 
In general, we introduce the Ricci coefficients as follows 
\beas
\chi_{AB} & = & g(D_A e_4, e_B)  \\ 
\underline{\chi}_{AB} & = & g(D_A e_3, e_B)  \\ 
\underline{\xi}_A & = & \frac{1}{2} g(D_3 e_3, e_A)  \\ 
\zeta_A & = & \frac{1}{2} g(D_3 e_4, e_A)  \\ 
\underline{\zeta}_A & = & \frac{1}{2} g(D_4 e_3, e_A)  \\ 
\nu & = & \frac{1}{2} g(D_4 e_4, e_3)  \\ 
\underline{\nu} & = & \frac{1}{2} g(D_3 e_3, e_4)  \\ 
\epsilon_A & = & \frac{1}{2} g(D_A e_4, e_3)
\eeas
\[\]

\subsection{Behavior at Future Null Infinity}
\label{future1}

We derive the following at future null infinity $\mathcal{I}^+$ for (A) spacetimes: 
\beas
\lim_{C_u, t \to \infty} r^3 \rho \ & = & \ P(u, \theta, \phi) \\ 
\bar{P}  \ & = & \ \bar{P} (u) \\ 
(P - \bar{P}) (u, \theta, \phi) & : & \mbox{ does not decay in $\mid u \mid$ as $\mid u \mid \to \infty$,} \\ 
& & \ \ \mbox{leading order term is dynamical, i.e. depends on $u$,} \\
& & \ \ \mbox{and also depends on the angles $\theta$, $\phi$} \\ 
\lim_{u \to + \infty} P(u, \theta, \phi)  \ & = & \ P^+ (\theta, \phi) 
\eeas
We see that $P = P(u, \theta, \phi)$ is a function on $R \times S^2$, and $P^+ = P^+ (\theta, \phi)$ is a function on $S^2$. 
Thus, in particular, as $u \to + \infty$, the quantity $P(u, \theta, \phi)$ tends to a function $P^+ (\theta, \phi)$ on $S^2$, not a constant.

Note that in \cite{sta} for (CK) spacetimes it is 
\beas
P - \bar{P} \ & = & \ O(\mid u \mid^{- \frac{1}{2}}) \\ 
\lim_{u \to + \infty} P  \ & = & \  P^+ \ = \  \lim_{u\to + \infty} \bar{P} \  =  \  \bar{P}^+ \ = \ -2M^+ \ = \ constant \\ 
\lim_{u \to - \infty} P \ & = & \ P^- \ = \  \lim_{u\to - \infty} \bar{P} \  =  \  \bar{P}^- \ = \ 0 
\eeas
Thus, the leading order term in $P$ as $u \to + \infty$ is a constant in the angles $\theta, \phi$ and cancels in $P - \bar{P}$. Here, $M$ denotes the Bondi mass and $M^+$ its limit for $u \to + \infty$, namely the ADM mass. 

\subsubsection{Memory}

{\bf Memory.} There is a natural contribution from $(P - \bar{P})$ to the gravitational wave memory effect. One may use the memory formulas from Christodoulou's paper \cite{chrmemory} or any of the present author's derivation of memory, for instance \cite{lydia3, lydia4}, but with the (A) data.

\subsection{Comparing (A) with (CK) Spacetimes}

\begin{tabbing}
\ \  \hspace{2cm} \= (A)  \hspace{6.9cm} \= (CK) \\  \\ 
$\rho$  \> $\lim_{C_u, t \to \infty} r^3 \rho  =   P(u, \theta, \phi)$ \> " \\ \\ 
$(P - \bar{P})$ \> does not decay in $\mid u \mid$ as $\mid u \mid \to \infty$,  \> $ O(\mid u \mid^{- \frac{1}{2}}) $ \\ 
\> tends to a non-zero function of $(\theta, \phi)$,  \> vanishes for $u \to + \infty$ \\ 
\> for $u \to + \infty$ \>   \\ \\ 
$\nlap \rho$  \> $O \ ( r^{- 4})$ \> $O \ ( r^{- 4} \tau_-^{- \frac{1}{2}})$ \\ \\ 
$\beta$ \> $o  (r^{- \frac{7}{2}})$ \> $o  (r^{- \frac{7}{2}})$ \\ \\ 
$\beta_3$ \> $O \ ( r^{- 4})$ \> $O \ ( r^{- 4} \tau_-^{- \frac{1}{2}})$ \\ \\ 
$\underline{\beta}_4$ \> $O \ ( r^{- 4})$ \> $O \ ( r^{- 4} \tau_-^{- \frac{1}{2}})$

\end{tabbing}

\[\]

\subsection{Energies and Control of the Curvature}
\label{energiescurvature}

In \cite{lydia5} we use energies to control the curvature components that are between the borderline case of \cite{lydia1, lydia2} and the strongly asymptotically flat situation of \cite{sta}. Whereas already in the former proof the present author had to work directly with the Bianchi equations in connection with the ``rougher" energies related to fewer vectorfields and with very slow decay of the data (borderline decay), the latter result by Christodoulou and Klainerman made use of the rotational vectorfields which gained them extra decay.

Whereas in \cite{sta}, the vectorfields $T, S, K, \bar{K}, O_i$ play crucial roles, in \cite{lydia1, lydia2}, in \cite{lydia5} and in the present situation, only the first four of these vectorfields come into play, no rotational vectorfields $O_i$ are at hand. Given the decay of the data in the present article, and the vectorfields in connection with the energies, more precisely the lack of symmetry (in particular when compared to the situation in \cite{sta}), 
we rely directly on the Bianchi equations. 
The setting investigated in this article requires different energies to be introduced and controlled. The estimates close at the optimal level due to the balancing between the energy estimates and the Bianchi equations and using extra information from the structure equations. In particular, to establish the optimal behavior for the curvature term $\rho$ and its derivatives, as well as for $\beta_3$ and $\underline{\beta}_4$ extra structures of these equations are used.

\section{Bianchi Equations and Structure Equations}

\subsection{Bianchi Equations}
\label{Bianchi}

The Bianchi equations read as follows. 

With respect to the null foliation, the Bianchi equations take the form: 
\bea
\Dlap_3 \alpha \ + \ \frac{1}{2} tr \underline{\chi} \alpha \ & = & \ 
- \ 2 \ \cDlap^*_2  \beta \ - \ 
3 \hat{\chi} \rho - 3 ^*\hat{\chi} \sigma  \ + \ 
2 \underline{\nu} \alpha + (\epsilon + 4 \zeta) \hat{\otimes} \beta 
\label{Bianchialpha3}  \\ \nonumber \\ 
\Dlap_3  \beta \ + \ tr \underline{\chi} \beta \ & = & \ 
\cDlap^*_1 (- \rho, \sigma) + 2 \hat{\chi} \underline{\beta} \ + \ 
 3 \zeta \rho + 3 ^*\zeta \sigma 
+  \underline{\nu} \beta + \underline{\xi} \alpha 
  \label{Bianchitubeta3}  \\ \nonumber \\ 
\Dlap_4 \beta \ + \ 2 tr \chi \beta  \ & = & \ 
\dlap \alpha \ - \ 
\nu \beta \ + \ 
(2 \epsilon + \zeta) \alpha 
  \label{Bianchitubeta4}  \\ \nonumber \\ 
\Dlap_3  \rho \ + \ \frac{3}{2} tr \underline{\chi} \rho \ & = & \ 
 - \dlap \underline{\beta} 
 - \frac{1}{2} \hat{\chi} \underline{\alpha} \ + \ 
 ( \epsilon  - \zeta) \underline{\beta} \ + \ 
2 \underline{\xi} \beta 
 \label{Bianchiturho3}  \\ \nonumber \\ 
\Dlap_4  \rho \ + \ \frac{3}{2} tr \chi \rho \ & = & \ 
\dlap \beta -  \frac{1}{2} \underline{\hat{\chi}} \alpha \ + \ 
\epsilon \beta + 2 \underline{\zeta} \beta 
  \label{Bianchiturho4}  \\ \nonumber \\ 
\Dlap_3 \sigma \ + \ \frac{3}{2} tr \underline{\chi} \sigma \ & = & \ 
 - \clap \ \ \underline{\beta}  - \frac{1}{2} \hat{\chi} ^*\underline{\alpha}  + 
\epsilon ^*\underline{\beta} - 2 \zeta ^*\underline{\beta} 
 - 2 \underline{\xi} ^*\beta  
  \label{Bianchitusigma3}  \\ \nonumber \\ 
\Dlap_4 \sigma \ + \ \frac{3}{2} tr \chi \sigma \ & = & \ 
- \clap \ \ \beta + \frac{1}{2} \underline{\hat{\chi}} ^* \alpha  \ + \ 
- \epsilon ^*\beta - 2 \underline{\zeta} ^*\beta 
  \label{Bianchitusigma4} \\ \nonumber \\ 
\Dlap_3  \underline{\beta} \ + \ 2 tr \underline{\chi} \underline{\beta} \ & = & \ 
- \dlap \underline{\alpha} - 
 (\zeta - 2 \epsilon) \underline{\alpha} - 
 \underline{\nu} \underline{\beta} - 
3 \underline{\xi} \rho + 3 ^*\underline{\xi} \sigma 
  \label{Bianchituunderlinebeta3}  \\ \nonumber \\
\Dlap_4  \underline{\beta} \ + \ tr \chi \underline{\beta} \ & = & \ 
\cDlap^*_1 (\rho, \sigma) + \nu \underline{\beta} + 
2 \underline{\hat{\chi}} \beta  - 
3(\underline{\zeta} \rho - ^*\underline{\zeta}\sigma) 
 \label{Bianchituunderlinebeta4}  \\ \nonumber \\ 
\Dlap_4 \underline{\alpha} \ + \ \frac{1}{2} tr \chi \underline{\alpha} \ & = & \ 
2 \cDlap^*_2 \underline{\beta} + 
2 \nu \underline{\alpha}  + 
(\epsilon - 4\underline{\zeta}) \hat{\otimes} \underline{\beta} 
- 3( \underline{\hat{\chi}} \rho - ^*\underline{\hat{\chi}} \sigma )
 \label{Bianchituunderlinealpha4}  \\ \nonumber 
 \eea

\subsection{Structure Equations}
\label{structure}

\bea
\dlap \hat{\underline{\chi}} & = & \underline{\beta} + \hat{\underline{\chi}} \cdot \zeta 
+ \frac{1}{2} ( \nlap tr \underline{\chi} - tr \underline{\chi} \zeta ) \ = \   \underline{\beta} + l.o.t. \label{dchibund1} \\ 
\dlap \hat{\chi} & = & - \beta - \hat{\chi} \cdot \zeta + \frac{1}{2} ( \nlap tr \chi + tr \chi \zeta )  \label{dchib1}
\eea
Recall that $\zeta$ is the torsion-one-form, introduced above in subsection \ref{Riccicoefficients1}.

The shears are related to each other by the equation 
\be \label{chihat}
\frac{\partial}{\partial u} \hat{\chi} \ = \ \frac{1}{4} tr \chi \cdot \hat{\underline{\chi}} + l.o.t. 
\ee
Also, it is 
\be \label{uchihat}
\frac{\partial}{\partial u} \underline{\hat{\chi}} \ = \ \frac{1}{2} \underline{\alpha} + l.o.t. 
\ee
\[\]

\section{Limits at Future Null Infinity $\mathcal{I^+}$} 
\label{future}

As a direct consequence from the behavior established in \cite{lydia5}, see section \ref{main1} above, we obtain the following theorem. 

\begin{The} \label{nummer1}
For (A) spacetimes, the normalized curvature components $r\underline{\alpha }$, $r^{2}\underline{\beta } $, $r^3 \rho $, $r^3 \sigma $ have limits on $C_u$ as $t\rightarrow \infty $: 
\begin{eqnarray*}
\lim_{C_{u},t\rightarrow \infty }r\underline{\alpha }
&=&A\left( u,\cdot \right) ,\, \ \ \ \ \ \ \ \ \ \ \ \
\lim_{C_{u},t\rightarrow \infty }\,r^{2}\underline{\beta }
= \underline{B}\left( u,\cdot \right) \ , \\ 
\lim_{C_{u},t\rightarrow \infty} r^3 \rho & = & P(u, \cdot) \ ,  \ \ \ \ \ \ \ \ \ \ \ \ 
\lim_{C_{u},t\rightarrow \infty} r^3 \sigma  =  Q(u, \cdot) 
\end{eqnarray*}
where the limits are on $S^{2}$ and depend on $u$. These limits satisfy 
\begin{eqnarray*}
\left| A \left( u,\cdot \right) \right| &\leq &C\left( 1+\left| u\right|
\right) ^{-5/2}\, \, \ \ \ \ \ \ \ \ \ \ \ \ \left| \underline{B}\left( u,\cdot
\right) \right| \leq C\left( 1+\left| u\right| \right) ^{-3/2}   \ \ \\ 
\left| Q \left( u,\cdot \right) \right| &\leq &C\left( 1+\left| u\right|
\right) ^{-1/2}\,
\end{eqnarray*}
whereas $P (u, \cdot)$, $(P (u, \cdot) - \bar{P} (u))$ do not decay in $|u|$. 
\end{The}
Moreover, the following limits exist  
\bea \label{limXiThe1}
\lim_{C_{u},t\rightarrow \infty } r^2 \widehat{\chi } & =: & \Sigma (u, \cdot)  \label{limSigmaThe1}  \\ 
- \frac{1}{2} \lim_{C_{u},t\rightarrow \infty }r\widehat{\underline{\chi }} & = &  \lim_{C_{u},t\rightarrow \infty } r \hat{\eta} =: 
\Xi \left( u,\cdot \right) \label{limXiThe1} 
\eea 
Further, it follows from (\ref{uchihat}), respectively from (\ref{chihat}) and from (\ref{dchibund1}) that 
\begin{eqnarray}
\frac{\partial \Xi }{\partial u} &=&-\frac{1}{4}A  \label{1XiAT1} \\ 
\frac{\partial \Sigma }{\partial u} &=& - \Xi \label{1SigmaT1}  \\ 
\underline{B} &=& - 2 \dlap \Xi \label{Ldchibund1}
\end{eqnarray}

Another consequence is the behavior 
\be \label{Xibsu}
\left| \Xi  \left( u,\cdot \right) \right| \leq C\left( 1+\left| u \right|
\right) ^{-3/2} 
\ee

We introduce $F/ 4 \pi$, the energy radiated away per unit angle in a given direction, where 
\be \label{Fenergy}
F( \cdot ) = \frac{1}{2} \int_{- \infty}^{+ \infty} | \Xi (u, \cdot ) |^2 du  \ . 
\ee
And we write $F(u)$ for 
\be \label{Fenergyu}
F(u, \cdot )  = \frac{1}{2} \int_{u}^{+ \infty} | \Xi (u', \cdot ) |^2 du'  \ . 
\ee
\[\]

{\scshape Behaviors of $\rho$ and $P$ versus $\sigma$ and $Q$:}  
Whereas the curvature terms $\rho$, $(\rho - \overline{\rho})$ and their limits $P (u, \cdot)$, $(P (u, \cdot) - \bar{P} (u))$ do not decay in $|u|$ as $|u| \to \infty$, the curvature terms $\sigma$, $(\sigma - \overline{\sigma})$ and the corresponding limits $Q (u, \cdot)$, $(Q (u, \cdot) - \bar{Q} (u))$ do. 
To this end, we observe the following: 
Whereas in the Christodoulou-Klainerman proof \cite{sta} for (CK) spacetimes the decay of these terms follows from the energy estimates using rotational vectorfields, the latter are not available in the present non-symmetric case. Instead the decay of 
$\sigma$ follows from its relation to $k$. In particular, we have for Einstein-vacuum spacetimes on each spacelike hypersurface $H_t$ 
\be \label{ksigma}
(curl \ k)_{lm} \ = \ H_{lm}
\ee
where $H$ is the magnetic part of the Weyl curvature as defined in (\ref{Wma1}), and $k$ behaves as in (\ref{initialdk1}). 
Decomposing this equation into parts that are tangential to and orthogonal to the surfaces $S_{t,u}$, the $NN$-component of equation (\ref{ksigma}) reads 
\be \label{ksigma2}
(curl \ k)_{NN} \ = \ \sigma \ . 
\ee
Consider the initial data for (A) spacetimes, and observe in (\ref{initialdk1}) how $k$ falls off. 
As shown in \cite{lydia5}, this behavior is preserved under the evolution by the Einstein equations. Thus, from (\ref{initialdk1}) and 
(\ref{ksigma2}) it follows that $\sigma$ and therefore also its limit $Q$ decay in $|u|$ as $|u| \to \infty$.

\section{Investigating $\rho$ via Bianchi Equations and Structure Equations}

Recall that we have $T = \frac{1}{\Phi}  \frac{\partial}{\partial t}$. Let $T = E_0$ 
as well as $(E_1, E_2, E_3)$ an orthonormal frame field for $H_t$. Thus we have the frame field $(E_0, E_1, E_2, E_3)$ for the spacetime $M$. 

We shall make use of the Bianchi equations and the structure equations at various levels. The Bianchi identities (themselves as well as in contracted forms) are used widely in the spacetime $(M, g)$ but also on the spacelike hyersurfaces $(H_t, \bar{g}_{ij}(t), k_{ij}(t))$. 

Equations (\ref{k1})-(\ref{intmaxequs7}) follow easily. 
We recall that the Codazzi equations read 
\be
\bar{\nabla}_i k_{jm} - \bar{\nabla}_j k_{im} = R_{m0ij}
\ee
and the Gauss equations take the form 
\be
\bar{R}_{imjn} + k_{ij}k_{mn} - k_{in} k_{mj} = R_{imjn} \ \ . 
\ee
By overline we denote the corresponding induced quantities on $H_t$. Thus, $\bar{R}_{imjn}$ are the components of the curvature tensor of $(H_t, \bar{g}(t))$.

Now, we consider the initial hypersurface $(H_0, \bar{g}_{ij}, k_{ij})$. 
Taking the trace of the Gauss equations yields 
\be \label{Gausstr*1}
\bar{R}_{ij} + tr k \  k_{ij} - k_{im} k^m_{\ j} = R_{ij} + R_{i0j0} \ \ . 
\ee
Thus, from (\ref{Gausstr*1}) using the EV equations (\ref{EV}) and the constraint equation (\ref{k1}) follows (\ref{k4}) 
\[
\bar{R}_{ij} \  =  \ k_{im} k^m_j \ + \ E_{ij}  \ . 
\]
Using the constraint equation (\ref{intmaxequs3}) together with the assumptions 
(\ref{initialdk1}) on the second fundamental form $k$ on $H_0$, it follows from (\ref{k4}) that 
\bea 
\bar{R}_{AB} & = & \frac{1}{4} \underline{\alpha}_{AB} + \frac{1}{4} \alpha_{AB} - \frac{1}{2} \rho \delta_{AB} + l.o.t.  \label{Riccistruct*1} \\ 
\bar{R}_{NN} & = & \rho + l.o.t.   \label{Riccistruct*2}  \\ 
\bar{R}_{AN} & = & - \frac{1}{2} ( \underline{\beta}_A  + \beta_A ) + l.o.t.            \label{Riccistruct*3} 
\eea

We obtain from the contracted Bianchi identities 
\be \label{barR1}
\bar{\nabla}_m \bar{R}^m_{ \ l} = \frac{1}{2} \bar{\nabla}_l \bar{R} \ , 
\ee
and from (\ref{intRbar1}) that 
\be \label{barR2}
\bar{\nabla}_m \bar{R}^m_{ \ l} = \frac{1}{2} \bar{\nabla}_l \bar{R} = o(r^{-6}) \ . 
\ee
Therefore, up to leading order it is 
\be \label{barR3}
\bar{\nabla}_m \bar{R}^m_{ \ l} =  0 \ . 
\ee
From (\ref{barR3}) and the components (\ref{Riccistruct*1})-(\ref{Riccistruct*3}) we deduce the equations 
\bea
tr \overline{Ric} \ & = & \ \bar{R} \ = \ 
\overline{R}_{ \ N}^N  + \overline{R}_{ \ A}^A \\ 
 \ & = & \ 0_{\mbox{\tiny leading order curvature terms}} + l.o.t. \ = \ o(r^{-5})
\eea

Therefore, we have up to leading order 
\be  \label{nRbar1}
\bar{\nabla}^i \bar{R}_{iB} \ = \ \bar{\nabla}^A \bar{R}_{AB}  + \bar{\nabla}^N \bar{R}_{NB} \ = \ 0 
\ee
with 
\be  \label{nRbar2}
\bar{\nabla}^A \bar{R}_{AB}  \ = \ \frac{1}{4} \bar{\nabla}^A  \underline{\alpha}_{AB} + \frac{1}{4} \bar{\nabla}^A  \alpha_{AB} 
- \frac{1}{2} \bar{\nabla}^A  \rho \cdot \delta_{AB}  
\ee
and 
\be  \label{nRbar3}
 \bar{\nabla}^N \bar{R}_{NB} \ = \ 
 -  \frac{1}{2} \bar{\nabla}^N \underline{\beta}_B  -  \frac{1}{2} \bar{\nabla}^N \beta_B \ . 
\ee
\[\]

\subsection{Limits at Spacelike Infinity}
\label{lsinf1}

In the following, lower order terms (l.o.t.) will be $o(r^{- \frac{11}{2}})$ or with stronger fall-off. 
\begin{The} \label{haupt1}
The following equation holds: 
\be \label{alar1}
\frac{1}{2} \nlap^A \alpha_{AB} + \frac{1}{2} \nlap^A \underline{\alpha}_{AB} -  \nlap_B \rho 
- ( \nlap^N  \underline{\beta}_B + \nlap^N \beta_B ) 
+ F_{\beta, \underline{\beta}} 
+ G (\Phi, \Psi)
 \ = \ 0 \ , 
\ee 
where 
$F_{\beta, \underline{\beta}}$ denotes terms of the types $\{ \mbox{constant} \cdot \frac{1}{r} \cdot \beta_B \}$ and $\{ \mbox{constant} \cdot \frac{1}{r} \cdot \underline{\beta}_B \}$; and $G (\Phi, \Psi)$ terms of the types $\{ \Phi \cdot \Psi \}$ with 
$\Psi$ denoting a Weyl curvature component, and 
$\Phi$ denoting a rotation coefficient except for the highest order term in $tr \chi$ and the highest order term in $tr \underline{\chi}$.
\end{The}
{\itshape Remark:} The $F_{\beta, \underline{\beta}}$ result from the highest order terms of products between $tr \chi$ and $tr \underline{\chi}$ on the one side and $\beta$ and $\underline{\beta}$ on the other side. All the other terms of these products are of lower order and are absorbed in $G (\Phi, \Psi)$.

Next, we consider $H_0$, thus $t=0$. By $S_r$ we denote surfaces in $H_0$ that are diffeomorphic to $S^2$. Let $\xi^B$ denote a conformal Killing vector field of $S_r$. The results for $S_r$ hold for any topological spheres. In our situation, we shall work with the $S_r$ that are induced by the intersections $H_0 \cap C_u$ at $t=0$, where $u = r = \underline{u}$.

\begin{The}  \label{haupt2}
(a) The integral of $F_{\beta_B, \underline{\beta}_B} \cdot \xi^B$ on $S_r$ is zero $\forall \xi^B$, 
\[
\int_{S_r} F_{\beta_B, \underline{\beta}_B} \cdot \xi^B = 0 \ . 
\]
(b) The integral of $G (\Phi, \Psi) \cdot \xi^B$ on $S_r$ is lower order $\forall \xi^B$, 
\[
\int_{S_r}  G (\Phi, \Psi) \cdot \xi^B = o(r^{- 2}) \ , 
\]
%
%
and 
\[
\lim_{r \to \infty} \int_{S_r}  G (\Phi, \Psi) \cdot \xi^B = 0 \ . 
\]
\end{The}

From theorems \ref{haupt1} and \ref{haupt2} the next result immediately follows: 
\begin{The}  \label{haupt3} 
Multiply equation (\ref{alar1}) by $\xi^B$ and integrate on $S_r$ to obtain up to leading order: 
\be \label{alar2}
 \frac{1}{2} \int_{S_r}  \nlap^A \alpha_{AB} \cdot \xi^B + \frac{1}{2} \int_{S_r} \nlap^A \underline{\alpha}_{AB}  \cdot \xi^B  -  \int_{S_r} \nlap_B \rho  \cdot \xi^B 
 \ = \ 0 \ \ \ \ \forall \ \xi^B \ . 
\ee
\end{The}

\begin{The}  \label{haupt4} 
We have to leading order 
\be \label{alar3}
  \int_{S_r} \nlap_B \rho \cdot \xi^B 
 \ = \ 0 \ \ \ \ \forall \ \xi^B \ . 
\ee
\end{The}
{\bf Proof:} This follows from the following standard result for the divergence operator on any $S_r$ diffeomorphic to the standard sphere $S^2$: The divergence operator acting on a $2$-covariant, symmetric, traceless tensor $t$, thus its image is the $1$-form $\dlap t$, is injective, and the range consists of all $L^2$-integrable $1$-forms on $S_r$ that are $L^2$-orthogonal to the Lie algebra of the conformal group of $S_r$. 
Consequently, the first two integrals in (\ref{alar2}) are zero, which proves the theorem.

Next, integrate (\ref{alar3}) by parts to obtain 
\be \label{alar4}
 \int_{S_r} \rho \cdot  \nlap_B \xi^B 
 \ = \ 0 \  \ \ \ \forall \  \xi^B \ . 
\ee

We want to study limits at infinity, therefore take the limit 
\bea
\lim_{H_0, r \to \infty} r \int_{S_r} \rho \cdot  \nlap_B \xi^B 
 \ & = & \ 0  \label{alar4} \\ 
 \ & = & \ 
 \int_{S^2} P_{H_0} (\theta, \phi)  \cdot  \nlap_B \xi^B \ \ \ \  \forall \  \xi^B  \ ,  \label{alar5}
\eea
where $P_{H_0} (\theta, \phi)$ denotes the limit of $r^3 \rho$ at spacelike infinity. 

We recall the well-known fact about the Laplace operator on the sphere $S^2$ and for every conformal Killing vectorfield $\xi^B$ 
\be \label{Laplace1}
\slap \dlap \xi + 2 \dlap \xi = 0 \ . 
\ee
With other words, $\dlap \xi$ for evey $\xi^B$ belongs to the first eigenspace, thus $l = 1$, for $\slap$. 
Moreover, each $l = 1$ spherical harmonic is the divergence of a unique conformal Killing vectorfield $\xi^B$ on $S^2$. 

Therefore, we conclude the following theorem.
\begin{The}  \label{haupt5} 
$P_{H_0} (\theta, \phi)$ does not have any $l=1$ mode. 
\end{The}

Next, we study the behavior at future null infinity $\mathcal{I}^+$. 
Consider 
\be \label{obv1}
\lim_{C_u, r \to \infty} r \int_{S_r} \rho \cdot \nlap_B \xi^B \ = \ 
\int_{S^2} P(u, \theta, \phi) \cdot \nlap_B \xi^B  \ =: \ \mA (u) \ . 
\ee
Obviously, it is 
\be \label{obv2}
\lim_{u \to \infty} \int_{S^2} P(u, \theta, \phi) \cdot \nlap_B \xi^B \ = \ 
\int_{S^2} P^+ (\theta, \phi) \cdot \nlap_B \xi^B \ . 
\ee

Now, whereas in this general setting, the limits for $\rho$ taken in the spacelike slice $H_0$ for $r \to \infty$, respectively the limit at $\mathcal{I}^+$ for $u \to \infty$, namely 
$P_{H_0} (\theta, \phi)$ and $P^+ (\theta, \phi)$ do not coincide, the corresponding integrals over spheres do. The reason for the latter is that these limiting integrals do not depend on the exhaustion. See Appendix \ref{A}. 
Thus, we have 
\be
P_{H_0} (\theta, \phi) \ \neq \ P^+ (\theta, \phi) 
\ee
however 
\be
\int_{S^2} P_{H_0} (\theta, \phi) \cdot \dlap \xi  \ = \ \int_{S^2} P^+ (\theta, \phi)  \cdot \dlap \xi  \ . 
\ee
From theorem \ref{haupt5} it follows that 
\be
\int_{S^2} P^+ (\theta, \phi) \cdot \nlap_B \xi^B \ = \ 0  \ \ \ \  \forall \  \xi^B \label{alar6} 
\ee
proving the next theorem. 
\begin{The}  \label{haupt6} 
$P^+ (\theta, \phi)$ does not have any $l=1$ mode. 
\end{The}

The next question is what happens at $\mathcal{I}^+$ for general $u$ as well as for $u \to - \infty$. 
We shall prove the next theorem. 
\begin{The} \label{haupt7} 
\noindent 
\begin{itemize} 
\item[(a)] Regarding $P$: For finite $u$ we have 
\be \label{a*1}
\int_{S^2} P(u) \cdot \nlap_A \xi^A \ = \ 
+ 2 \int_{S^2} F(u) \cdot \nlap_A \xi^A \ - \ 
\int_{S^2} \Sigma(u) \Xi(u) \cdot \nlap_A \xi^A \ . 
\ee
\item[(b)] Regarding $P$: For $u \to - \infty$ it is 
\be \label{b*1} 
\int_{S^2} P^- \cdot \nlap_A \xi^A     \ = \  + 2 \int_{S^2} F \cdot \nlap_A \xi^A \ . 
\ee
\item[(c)] Regarding $Q$: For $u \to - \infty$, respectively $u \to + \infty$, it is 
\be \label{c*1} 
Q^- = Q^+ = 0 \ \ , \ \ \mbox{therefore trivially} \ \ Q^-_{l=1} = Q^+_{l=1} = 0
\ee
\item[(d)] Regarding $Q$: For finite $u$ we have 
\be \label{d*1}
\int_{S^2} Q(u) \cdot \nlap_A \xi^A \ = \ 
 \ - \ 
\int_{S^2} (\Sigma(u) \wedge \Xi(u)) \cdot \nlap_A \xi^A \ . 
\ee
\end{itemize}
\end{The}
{\bf Proof:} 
First, we derive a Hodge system of equations on $S^2$. Then we apply Hodge theory (see appendix \ref{Hodge}) to prove the statements. 

{\itshape Equations for $\rho$ and $P$:} 
Recall the Bianchi equation (\ref{Bianchiturho3}) for $\Dlap_3 \rho$ 
\[
\Dlap_3  \rho \ + \ \frac{3}{2} tr \underline{\chi} \rho \  =  \ 
 - \dlap \underline{\beta} 
 - \frac{1}{2} \hat{\chi} \underline{\alpha} \ + \ 
 ( \epsilon  - \zeta) \underline{\beta} \ + \ 
2 \underline{\xi} \beta 
\ = \ - \dlap \underline{\beta} 
 - \frac{1}{2} \hat{\chi} \underline{\alpha} \ + \ l.o.t. 
\] 
which we write as 
\be \label{Brho3*1}
\rho_3 \  =  \ 
 - \dlap \underline{\beta} 
- \frac{\partial}{\partial u} (\hat{\chi} \cdot \hat{\underline{\chi}})  + \frac{1}{4} tr \chi  |\hat{\underline{\chi}}|^2  + \ l.o.t. 
\ee
where $\rho_3 : = \Dlap_3  \rho \ + \ \frac{3}{2} tr \underline{\chi} \rho$, using 
(\ref{chihat}) and (\ref{uchihat}). 
Multiply equation (\ref{Brho3*1}) by $r^3$ and take the limit along $C_u$ as $t \to \infty$ to obtain 
\be  \label{Brho3*2} 
- 2 \frac{\partial}{\partial u} P \  =  \ 
- \dlap \underline{B} + 2 \frac{\partial}{\partial u} ( \Sigma \cdot \Xi ) + 2 | \Xi |^2  \ + \ l.o.t. 
\ee
Using (\ref{Ldchibund1}) as well as (\ref{1SigmaT1}) and integrating with respect to $u$ yields 
\bea
(P (u) - P^+) \ & = & \ \dlap \dlap (\Sigma (u) - \Sigma^+) 
+ \int_{u}^{+ \infty} | \Xi (u') |^2 du' \nonumber \\ 
\ & & \ - (  \Sigma (u) \cdot \Xi (u) -  \underbrace { \Sigma^+ \cdot \Xi^+}_{= 0}  )   \nonumber \\ 
 \ & = & \ 
 \dlap \dlap (\Sigma (u) - \Sigma^+) 
+ \int_{u}^{+ \infty} | \Xi (u') |^2 du' 
- \Sigma (u) \cdot \Xi (u)   \label{Pu1} \\ 
(P^- - P^+) \ & = & \ \dlap \dlap (\Sigma^- - \Sigma^+) 
+ \int_{- \infty}^{+ \infty} | \Xi (u) |^2 du  \nonumber \\ 
\ & & \  - ( \underbrace{ \Sigma^- \cdot \Xi^- }_{= 0} - \underbrace{ \Sigma^+ \cdot \Xi^+ }_{= 0}  )  \nonumber \\ 
 \ & = & \ \dlap \dlap (\Sigma^- - \Sigma^+) 
+ \int_{- \infty}^{+ \infty} | \Xi (u) |^2 du   \label{Pall1}  \ . 
\eea
Thus, we have 
\bea
(P (u) - P^+) \ & = & \ \dlap \dlap (\Sigma (u) - \Sigma^+) 
+ 2 F(u) 
- \Sigma (u) \cdot \Xi (u) \label{Pu2} \\ 
(P^- - P^+) \ & = & \ \dlap \dlap (\Sigma^- - \Sigma^+) 
+ 2 F  \ .  \label{Pall2}
\eea 

{\itshape Equations for $\sigma$ and $Q$:} 
Next, we consider the Bianchi equation (\ref{Bianchitusigma3})
\[
\Dlap_3 \sigma \ + \ \frac{3}{2} tr \underline{\chi} \sigma \  =  \ 
 - \clap \ \ \underline{\beta}  - \frac{1}{2} \hat{\chi} ^*\underline{\alpha}  + 
\epsilon ^*\underline{\beta} - 2 \zeta ^*\underline{\beta} 
 - 2 \underline{\xi} ^*\beta  
 \]
which we write as 
\be \label{Bsigmaauf1*}
\sigma_3 \  =  \ 
  - \clap \ \ \underline{\beta} 
- \frac{\partial}{\partial u} (\hat{\chi} \wedge \hat{\underline{\chi}}) 
 + \ l.o.t. 
\ee
where $\sigma_3 : = \Dlap_3  \sigma \ + \ \frac{3}{2} tr \underline{\chi} \sigma$, using 
(\ref{chihat}) and (\ref{uchihat}). 
Multiply (\ref{Bsigmaauf1*}) by $r^3$ and take the limit on $C_u$ as $r \to \infty$ to obtain 
\be 
Q_3 = - \clap  \ \ \underline{B}  + 2 \frac{\partial}{\partial u} (\Sigma \wedge \Xi)
\ee
that reads 
\be \label{BSigmaauf*2*}
 \frac{\partial Q}{\partial u}  \ = \ \frac{1}{2}  \clap  \ \ \underline{B}  -  \frac{\partial}{\partial u} (\Sigma \wedge \Xi) \ . 
\ee
We infer 
\bea
Q(u) - Q^+ \ & = & \ \clap \ \ \dlap (\Sigma(u) - \Sigma^+)  - \Sigma(u) \wedge \Xi(u) \label{Qlimitsu*1*} \\ 
Q^- - Q^+ \ & = & \  \ \clap \ \ \dlap (\Sigma^- - \Sigma^+)  \label{Qlimits*1*}
\eea

At this point, we make use of the Hodge theory from appendix \ref{Hodge}. 

From above we know that $Q = O(|u|^{- \frac{1}{2}})$. Note that the mean value $\bar{Q}$ of $Q$ on $S^2$ obeys $\bar{Q} = O(|u|^{- \frac{3}{2}})$. However, there is no power law decrease (nor increase) for the corresponding mean value $\bar{P}$ of $P$. 

{\scshape Derivation of equation (\ref{b*1})}: 
Consider the equation (\ref{Pall2}) and the independently derived equation (\ref{Qlimits*1*}). For 
\be  \label{Z1*}
\mathcal{Z}  = \dlap (\Sigma^- - \Sigma^+) 
\ee 
we derive the Hodge system 
\bea
\dlap \mathcal{Z}  & = &  (P - \bar{P})^- - (P - \bar{P})^+ - 2 (F - \bar{F})  \ ,   \label{Zauf2*} \\ 
\clap \ \ \mathcal{Z}  & = & 0  \ . \label{Zauf3*} 
\eea
On $S^2$, define the function $\varphi$ to be the solution of vanishing mean of the equation 
\be \label{Zauf4*}
\slap \varphi = (P - \bar{P})^- - (P - \bar{P})^+ - 2 (F - \bar{F}) \ .  
\ee
Then we have 
\be \label{Zauf1*}
\mathcal{Z} =  \nlap \varphi \ . 
\ee
Equations (\ref{Z1*}), (\ref{Zauf4*}), (\ref{Zauf1*}) determine the solution $(\Sigma^- - \Sigma^+)$ uniquely. 
The integrability condition is that $\varphi$ has vanishing projection onto the first eigenspace of $\slap$, namely $\varphi_{l=1} = 0$. 
Now, we compute the $l=1$ modes, noting that the elements of the $l=1$ eigenspace are orthogonal to the constants and taking into account theorem \ref{haupt6}. 
Then claim {\ref{b*1}} follows. 

{\scshape Equation (\ref{c*1})}: 
Equation (\ref{c*1}) trivially follows. 

{\scshape Derivation of equation (\ref{d*1})}: 
At this point, we consider equation (\ref{BSigmaauf*2*}): 
\[
\frac{\partial Q}{\partial u} \ =  \ \frac{1}{2}  \clap  \ \ \underline{B}   -  \frac{\partial}{\partial u} (\Sigma \wedge \Xi) \ .  
\] 
From theorem \ref{nummer1} we know that the first term on the right hand side is of order $O(|u|^{- \frac{3}{2}})$, whereas the last term on the right hand side is of order $O(|u|^{- \frac{5}{2}})$. Now, we look at the $l=1$ modes for $Q(u)$. 
Note that $(\clap \ \ \underline{B} )_{l=1} =0$. Therefore, the $l=1$ component of $Q$ will be of the form 
$Q_{l=1} (u) = a_i \xi^i + O(|u|^{- \frac{3}{2}})$, where $a_i$ are constants independent of $u$. 
Recall that 
 $Q = O(|u|^{- \frac{1}{2}})$ and $\bar{Q} = O(|u|^{- \frac{3}{2}})$; therefore it is also $(Q - \bar{Q}) = O(|u|^{- \frac{1}{2}})$. 
 It follows that $a_i = 0$, that is $Q_{l=1}(u) = O(|u|^{- \frac{3}{2}})$. 
 We conclude from (\ref{BSigmaauf*2*}) that 
\[
\int_{S^2} Q(u) \cdot \nlap_A \xi^A \ = \ - \int_{S^2} ( \Sigma (u) \wedge \Xi (u)) \cdot \nlap_A \xi^A  \ , 
\]
which is equation (\ref{d*1}).

{\scshape Derivation of equation (\ref{a*1})}: 
Consider the equation (\ref{Pu2}) and the independently derived equation (\ref{Qlimitsu*1*}). For 
\be \label{Zu1*}
\mathcal{Z}  = \dlap (\Sigma (u) - \Sigma^+)
\ee
we derive the Hodge system 
\bea
\dlap \mathcal{Z}  & = &  (\mathcal{P} - \bar{\mathcal{P}}) (u)  - (\mathcal{P} - \bar{\mathcal{P}})^+  - 2 (F(u) - \bar{F}(u))  \nonumber \\ 
& &  - ( \Sigma \cdot \Xi (u) -  \overline{\Sigma \cdot \Xi (u)})  \ ,   \label{Zuauf2*} \\ 
\clap \ \ \mathcal{Z}  & = & (\mathcal{Q} - \bar{\mathcal{Q}}) (u) + ( \Sigma \wedge \Xi (u) - \overline{ \Sigma \wedge \Xi (u)} )  \ . \label{Zuauf3*} 
\eea
On $S^2$, we define the function $\varphi$ to be the solution of vanishing mean of the equation 
\be \label{Zuauf4*}
\slap \varphi =  (\mathcal{P} - \bar{\mathcal{P}}) (u)  - (\mathcal{P} - \bar{\mathcal{P}})^+  - 2 (F(u) - \bar{F}(u))   - ( \Sigma \cdot \Xi (u) -  \overline{\Sigma \cdot \Xi (u)})   
\ee
and the function $\psi$ 
to be the solution of vanishing mean of the equation 
\be \label{Zuauf5*}
\slap \psi = (\mathcal{Q} - \bar{\mathcal{Q}}) (u) + ( \Sigma \wedge \Xi (u) - \overline{ \Sigma \wedge \Xi (u)} )  \ . 
\ee 
Then we have 
\be \label{Zuauf1*}
\mathcal{Z}  = \nlap \varphi + \nlap^{\perp} \psi  \ . 
\ee
It is 
\be \label{Sigmaintwirld0}
\int_{S^2}  \dlap (\Sigma (u) - \Sigma^+) \cdot \xi = 0 \ . 
\ee
That is 
\be
\int_{S^2} (\Sigma (u) - \Sigma^+) \cdot \dlap \xi = (\Sigma (u) - \Sigma^+)_{l=1}  = 0 \ . 
 \ee
Thus, we write with (\ref{Zu1*}) and (\ref{Zuauf1*}) 
\bea
\int_{S^2} Z \cdot \xi & = &  0   \\ 
 & = & 
  \int_{S^2} (\nlap \varphi) \cdot \xi \ + \  \int_{S^2} (\nlap^{\perp}  \psi) \cdot \xi  \\ 
   & = & 
 -  \int_{S^2} \varphi \cdot \dlap \xi -  \int_{S^2} \psi \cdot \underbrace{\clap \ \ \xi }_{ = - \dlap \xi^{\perp}} \\ 
  & = & 
 -  \int_{S^2} \varphi \cdot \dlap \xi +  \int_{S^2} \psi \cdot \dlap  \xi^{\perp} 
\eea
Using that 
\[
\slap \dlap \xi = - 2 \dlap \xi 
\]
we further compute 
\bea
 0 & = &  \int_{S^2} \varphi \cdot \slap \dlap \xi -  \int_{S^2} \psi \cdot \slap \dlap  \xi^{\perp}   \\ 
 & = & 
  \int_{S^2} \slap \varphi \cdot  \dlap \xi -  \int_{S^2} \slap \psi \cdot \dlap  \xi^{\perp}  \label{intalles}
\eea
We use that $\dlap \mathcal{Z} = \slap \varphi$ and $\clap \ \  \mathcal{Z} = \slap \psi$. See equations (\ref{Zuauf4*})-(\ref{Zuauf5*}). 
From the previous argument in the derivation of equation (\ref{d*1}) we obtain that the second integral in(\ref{intalles}) is zero. 
Then from the first integral in (\ref{intalles}), using (\ref{Zuauf4*}), we compute the $l=1$ modes, noting that the elements of the $l=1$ eigenspace are orthogonal to the constants and taking into account theorem \ref{haupt6}. 
Then claim (\ref{a*1}) follows. We conclude that $(P(u))_{l=1} = O(|u|^{- \frac{3}{2}})$. This concludes the proof of theorem \ref{haupt7}.  \\

\section{Angular Momentum at $\mathcal{I^+}$}
\label{angular}

In this section, we show that the angular momentum at future null infinity $\mathcal{I^+}$ is well defined for (A) spacetimes. 
Given the behavior of the curvature components at $\mathcal{I^+}$, it looks as if there was not enough decay for the angular momentum at $\mathcal{I^+}$ to be well defined. However, the corresponding $l=1$ modes of the crucial components, that enter the definition, behave better and in particular are integrable in $u$. 

We recall the classical definition of angular momentum at $\mathcal{I^+}$: 
\be \label{angm1}
J^k \ : = \ \int_{S^2} \epsilon^{AB} \nabla_B \tilde{X}^k  ( N_A - \frac{1}{4} C_A^{ \ D} \nabla^B C_{DB} ) \ \ , \ \ k = 1, 2, 3. 
\ee
Most physics literature work with Bondi-Sachs coordinates. In (\ref{angm1}), 
$\tilde{X}^k$ for $k=1, 2, 3$ denote the standard coordinate functions in $\real^3$ restricted to $S^2$, 
$N_A$ is the angular momentum aspect, $C_{AB}$ the shear tensor and $\epsilon_{AB}$ the volume form of the standard round metric $\sigma_{AB}$ of $S^2$. Further, in the said notation, $N_{AB}$ is the news tensor and $m$ the mass aspect. 
In this notation, $\nabla_A$ in (\ref{angm1}) is the covariant derivative with respect to $\sigma_{AB}$. Raising, lowering indices and contraction happens via the metric $\sigma_{AB}$. See for instance \cite{draystreu1}, \cite{flanagannichols1}. See also \cite{chenkellerwangwangyau} for a discussion of classical angular momentum and their newly introduced definition of angular momentum in a different setting than studied here. 
Note that (\ref{angm1}) is also the definition in \cite{arizzi1} but using the Christodoulou-Klainerman notation \cite{sta}. 

While for (CK) spacetimes angular momentum $J^k$ at $\mathcal{I^+}$ is well defined and conserved \cite{arizzi1}, \cite{chenkellerwangwangyau}, this was not clear for (A) spacetimes. We are now going to prove that indeed this holds for the more general (A) spacetimes. 
\begin{The} \label{AMA1}
Let $(M,g)$ be an (A) spacetime. Then angular momentum $J^k$ at $\mathcal{I^+}$ is well defined and conserved. 
\end{The}
{\itshape Remark:} In order to prove this theorem, we shall explore the ingredients in (\ref{angm1}) and how they behave at $\mathcal{I^+}$ of (A) spacetimes. Moreover, we use the notation as in \cite{sta} and \cite{lydia1}, \cite{lydia2}. For the readers' convenience, we give the `translation' between these notations in Appendix \ref{B}. \\ 

{\bf Proof of Theorem \ref{AMA1}:} 
As in (A) spacetimes the quantity corresponding to $N_A$ in (\ref{angm1}) may not have a limit at future null infinity along $C_u$ or not have enough decay (that is the limit at future null infinity along $C_u$ of $\beta_A$ multiplied by corresponding weights in $r$), we investigate this now in detail. 
Use the Bianchi equation (\ref{Bianchitubeta3}) for $\Dlap_3 \beta$ 
\bea
\Dlap_3  \beta \ + \ tr \underline{\chi} \beta \ & = & \ 
\cDlap^*_1 (- \rho, \sigma) + 2 \hat{\chi} \underline{\beta} \ + \ 
 3 \zeta \rho + 3 ^*\zeta \sigma 
+  \underline{\nu} \beta + \underline{\xi} \alpha   \nonumber \\ 
\ & = & \ 
\nlap \rho + \epsilon_{AB} \nlap^B \sigma + 2 \hat{\chi} \underline{\beta} + l.o.t.  \label{Bbu3lot}
\eea
As the right hand side of (\ref{Bbu3lot}) obeys good decay behavior (see section \ref{main1}), we multiply it with $r^4$ and take the limit on a given $C_u$ as $r \to \infty$. Each of the components on the right hand side has a well-defined limit at $\mathcal{I^+}$. Therefore, it follows that the left hand side tends to a well-defined limit at $\mathcal{I^+}$. We call this limit $R(u, \theta, \phi)$: 
\[
\lim_{C_u, r \to \infty} r^4 ( \Dlap_3  \beta + tr \underline{\chi} \beta ) =: R (u, \theta, \phi) \ . 
\]
And from (\ref{Bbu3lot}) we obtain the limiting equation at $\mathcal{I^+}$ 
\be \label{RPQSLB}
R = \nlap P + \ ^* \nlap Q + 2 \Sigma \cdot \underline{B} \ . 
\ee
Note that equation (\ref{RPQSLB}) was derived in \cite{chrIV2000} and used in \cite{chenkellerwangwangyau} but for different spacetimes, namely in those papers the leading order term of $\rho$ behaves differently as explained above, moreover $\nlap \rho$ decays in $u$. That is not the case for (A) spacetimes, for the latter are more general, in particular, the leading order term of $\rho$ includes more general terms and $\nlap \rho$ does not fall off in $u$. Further, we point out the different behavior of this spacetime outlined in the first part of the present article. Whereas the geometric quantities of the manifolds investigated in \cite{chrIV2000} and \cite{chenkellerwangwangyau} either are directly controlled by the results of \cite{sta} or by a computation that can be related to quantities controlled by \cite{sta}, this is not the case for (A) spacetimes. The latter can be controlled by results of \cite{lydia5}. In particular, various geometric quantities show rougher behavior, and it is not clear if the integrant of the integral in (\ref{angm1}), involving quantities at $\mathcal{I^+}$, is defined in the (A) setting.  
In fact, the analogues of the estimates as in \cite{chrIV2000} or \cite{chenkellerwangwangyau} would not work here 
because of these reasons just explained above. Here, we take a different approach and show that the integrant of the integral in (\ref{angm1}) is well-defined for (A) spacetimes. 

In the following, for a function $f$ on $S^2$ we write the projection of $f$ on the sum of the zeroth and first eigenspaces of $\slap$ as 
$f_{[1]}$, and use $f_{l=0}$, $f_{l=1}$ for the corresponding projections on zeroth, respectively first eigenspace. Thus, 
$f_{[1]} = f_{l=0} + f_{l=1}$. Also, for a $1$-form $r_A = \nlap_A f + \epsilon_{AB} \nlap^B g$ write 
$r_{A [1]} = \nlap_A f_{l=1} + \epsilon_{AB} \nlap^B g_{l=1}$. 

Compute from (\ref{RPQSLB}), take the $l=1$ modes and integrate to obtain 
\bea
\int_{u_1}^{u_2} R_{[1]} du & = &
 \int_{u_1}^{u_2} \nlap P_{l=1} du  \int_{u_1}^{u_2}  \ ^* \nlap Q_{l=1}  du + 2 \int_{u_1}^{u_2} \large( \Sigma \cdot \underline{B}  \large)_{[1]} du   \label{aufo1}
\eea
Let 
\be
\tilde{R}_{[1]} \ := \  - \frac{1}{2} \int R_{[1]} (u) du \ .
\ee
Then (in our sign convention for $u$) it is 
\be
\tilde{R}_{[1]} (u) - \tilde{R}_{[1]}^+ \ = \  - \frac{1}{2}  \int_{+ \infty}^{u} R_{[1]} (u') du' 
\ee
Claim: In (\ref{aufo1}), each term on the right hand side (RHS) is integrable. This is straightforward for the last two terms. 

In view of the last term on the RHS of (\ref{aufo1}) we know from theorem \ref{nummer1} and the statements therafter that 
$\Sigma \cdot \underline{B} = O(|u|^{- \frac{3}{2}})$. 

For the second term on the RHS of (\ref{aufo1}) we know from theorem \ref{haupt7} that $Q_{l=1}(u) = O(|u|^{- \frac{3}{2}})$. 

The integrability for the first term on the RHS of (\ref{aufo1}) follows directly from our theorem \ref{haupt7} above that gives 
$P_{l=1}(u) = O(|u|^{- \frac{3}{2}})$. 
Note that this is a consequence of the $l=1$ modes behaving much better than $P$ itself. In particular, $\nlap P$ depending on $u$ but not decaying in $|u|$ may suggest on a first look that the integral diverges. However, using (\ref{Xibsu}) and (\ref{Fenergyu}) in theorem \ref{haupt7} shows immediately that $P_{l=1}$ is integrable in $u$. This concludes the proof of theorem \ref{AMA1}. \\ 

{\itshape A Formula in the Direction of a Conservation Law of Angular Momentum:} In a straightforward manner, we derive from the above the following {formula}: 
\bea \label{consam1*}
& & \lim_{u \to + \infty} \large( \tilde{R}_{[1]} (u) - ( \Sigma \cdot \dlap \Sigma )_{[1]} \large) - 
\lim_{u \to - \infty} \large( \tilde{R}_{[1]} (u) - ( \Sigma \cdot \dlap \Sigma )_{[1]} \large) 
\nonumber \\ 
& & \ \ = \ \  \frac{1}{2} \int_{- \infty}^{+ \infty} - \nlap_A  P_{l=1} - \nlap^* Q_{l=1} + 
2 \large(  \Xi \cdot \dlap \Sigma - \Sigma \cdot \dlap \Xi \large)_{[1]} \  du   \label{consam1*}
\eea
\[\]

\section{Weyl Curvature Behavior at Future Null Infinity $\mathcal{I}^+$}

\subsection{Peeling Stops} 
\label{peelingstops}

From the Bianchi equations (\ref{Bianchiturho3}) and (\ref{Bianchitusigma3}) we derive the limiting equations at future null infinity $\mathcal{I}^+$ for the limits $P$, respectively $Q$. They read 
\bea
\frac{\partial P}{\partial u} \  & = & \ \frac{1}{2} \dlap \underline{B}  -  \Sigma \cdot \frac{\partial \Xi}{\partial u}   \\ 
\frac{\partial Q}{\partial u} \ & = & \ \frac{1}{2}  \clap  \ \ \underline{B}  -  \Sigma \wedge \frac{\partial \Xi}{\partial u} 
\eea
We observe that $P(\theta, \phi, u)$ at highest order does not have any power law decrease nor increase in $u$ as $|u| \to \infty$, but it depends on $u$ and changes with $u$. Then by these equations it must hold that 
\be
\frac{\partial P}{\partial u} \ = \ o(|u|^{- \frac{3}{2}}) \ . 
\ee
If we assume more decay, then we could also have 
\be
\frac{\partial P}{\partial u} \ = \ O(|u|^{- 2}) \ . 
\ee

Recall equation (\ref{RPQSLB}) from above 
\[
R = \nlap P + \ ^* \nlap Q + 2 \Sigma \cdot \underline{B} \ . 
\]
We obtain for $uR$ a behavior like $r^{-4} |u|^{+ 1}$, and correspondingly also for $\beta$ a behavior like $r^{-4} |u|^{+ 1}$. Thus, $\beta$ has less decay and the leading order term is dynamical.  \\

\section{Conclusions}
\label{conclusions}

{\bf Limits for $P(u, \theta, \phi)$}: Whereas $\sigma \to 0$ as 
$|u| \to \infty$, this is not true for $\rho$. This has the following deep implications. 
The fact that there are no extra constraints, other than by the constraint equations, put on the part of the initial metric that is homogeneous of degree $-1$, means that the leading order $O(r^{-3})$ portion of $\rho$ is dynamical, that is it depends on $u$. 
Namely, the constraint equation (\ref{intmaxequs3}) and the conditions on the initial data give (\ref{intRbar1}) 
\[
\bar{R}  \ =  \  |k|^2 \ = \ o(r^{-5})  \ , 
\]
yielding the fact stated in (\ref{didjhij*1}) which reads 
\[
\partial_i \partial_j h_{ij} - \partial_j \partial_j h_{ii}  \ = \ 0  \ \ . 
\] 
Thus, the homogeneous of degree $-3$ part of $\bar{R}$ vanishes. 
See equations (\ref{intRbar1})-(\ref{didjhij*1}) in chapter \ref{set}. 

In general, this dynamical term of $\rho$ will take different limits at $\mathcal{I}^+$ when $u \to + \infty$, respectively $u \to - \infty$. These yield (\ref{Prholimit1}). 
\be \label{Prholimit1} 
P(u, \theta, \phi) \to P^+ (\theta, \phi)  \ \  \mbox{ as } \ \  u \to + \infty \ . 
\ee
In particular, the limit  $P^+ (\theta, \phi)$ is not a constant but rather a function on $S^2$. \\

\section*{Acknowledgments}

The author thanks the NSF, acknowledging support from the NSF Grants DMS-2204182 and DMS-1811819. \\

\appendix 

\section{Integrals at Spacelike Infinity} 
\label{A}

In this appendix, we show that: 
The following limiting integrals are independent from the exhaustion and therefore coincide 
\be 
\int_{S^2} P^+ (\theta, \phi) \cdot \nlap_A \xi^A \ = \ \int_{S^2} P_{H_0} (\theta, \phi) \cdot \nlap_A \xi^A   \ \ \ \  \forall \  \xi^A \label{appe1} \ . 
\ee

First, consider $H_0$. 
Let $B_n$ be domains in $H_0$ such that $B_{n+1} \supset B_n$ and $\cup_n B_n = H_0$ with $S_n = \partial B_n$. (Naturally the $S_n$ are $C^1$.) Then from (\ref{alar4}) - (\ref{alar5}) we know that 
\[
\lim_{H_0, n \to \infty} r \int_{S_n} \rho \cdot  \nlap_B \xi^B 
\ = \ 
 \int_{S^2} P_{H_0} (\theta, \phi)  \cdot  \nlap_B \xi^B \ \ \ \  \forall \  \xi^B 
 \]
Now, we consider two domains $B_2 \supset B_1$ in $H_0$ and such that $B_1$ contains the coordinate ball of radius $R$. 
We obtain 
\be
 \int_{S_2}    \rho \cdot  \nlap_B \xi^B     -  \int_{S_1}  \rho \cdot  \nlap_B \xi^B 
 \ = \ 
 \int_{B_2 \backslash B_1} \partial_i {\large{(}} \rho \cdot \nlap_B \xi^B {\large{)}} \ . 
\ee
From above we know that the integrand on the right hand side is $O(r^{-4})$. Therefore, we conclude that 
\be
 \int_{B_2 \backslash B_1} \partial_i {\large{(}} \rho \cdot \nlap_B \xi^B {\large{)}} \ \leq \ 
 C R^{-1} \ \to \ 0 \ \ \mbox{as} \ \ R \to \infty \ . 
\ee 
Thus the limit does not depend on the exhaustion. 

Next, consider $\mathcal{I}^+$. 
From (\ref{obv1}) - (\ref{obv2}) we know that 
\[
\lim_{C_u, r \to \infty} r \int_{S_r} \rho \cdot \nlap_B \xi^B \ = \ 
\int_{S^2} P(u, \theta, \phi) \cdot \nlap_B \xi^B  \ =: \ \mA(u) \ . 
\]
\[
\lim_{u \to \infty} \int_{S^2} P(u, \theta, \phi) \cdot \nlap_B \xi^B \ = \ 
\int_{S^2} P^+ (\theta, \phi) \cdot \nlap_B \xi^B \ . 
\] 
Let $u_2 > u_1$. Eventually, we will let both $u_1, u_2 \to + \infty$. 
A short computation shows that it is 
\be \label{appeAu1u2}
\mA (u_2) - \mA (u_1) \ = \ \int_{u_1}^{u_2} \int_{S^2} \frac{\partial}{\partial u} P \cdot \nlap_B \xi^B \ . 
\ee
And we know from above that 
$\frac{\partial}{\partial u} P = O(| u |^{- \frac{3}{2}})$ as $|u| \to \infty$. 
Thus, we conclude that 
\be \label{gleichen1}
\int_{u_1}^{u_2} \int_{S^2} \frac{\partial}{\partial u} P \cdot \nlap_B \xi^B \ \to \ 0  \ \ \mbox{as} \ \ u \to \infty \ . 
\ee
Thus the limit does not depend on the exhaustion either. 

Next, we keep $u_1$ fixed in (\ref{appeAu1u2}) and let $u_2 \to \infty$. 
Then in $H_0$ the surface $S_{0, u_2}$ tends to infinity. 
Thus, as the limit from (\ref{alar4}) - (\ref{alar5}) in $H_0$ does not depend on the foliation $\{ S_r \}$, we may pick the foliation given by 
$S_{0, u_2} = C_{u_2} \cap H_0$ as $u_2 \to \infty$. 
Then equation (\ref{appe1}) follows from equations (\ref{appeAu1u2}) and (\ref{gleichen1}). \\

\section{Notation}
\label{B}

We relate the Christodoulou-Klainerman notation to the Bondi-Sachs coordinate system. For a nice derivation of further components, see \cite{chenkellerwangwangyau}. But note that we use slightly different conventions in the current article than is used in the latter paper. 
In the following, the left hand side is given in the Christodoulou-Klainerman notation: 
\beas
B_A \ & = & \ - N_A \ \ = \ \ - I_A \\ 
\underline{B}_A \ & = & \ \nlap^B N_{AB} \\ 
\underline{A}_{AB} \ & = & \ - 2 \partial_u N_{AB} \\ 
\Sigma_{AB} \ & = & \ - \frac{1}{2} C_{AB} \\ 
 \Xi_{AB} \ & = & \ - \frac{1}{2}  N_{AB} \ . 
\eeas
\[\]

\section{Hodge Theory at Future Null Infinity}
\label{Hodge}

The following Hodge systems and results from Hodge theory are frequently used in this article. In particular, this is in connection with the equations (\ref{Pu2}) and (\ref{Qlimitsu*1*}), as well as (\ref{Pall2}) and (\ref{Qlimits*1*}). 

Let $Z$ be a sufficiently smooth vector field on $S^2$. There exist scalar fields $\varphi$ and $\psi$ such that 
\[
Z = \nlap \varphi + \nlap^{\perp} \psi  \ . 
\]
Then we have 
\[
\dlap Z = \slap \varphi \ \ \ \ , \ \ \ \  \clap \ \ Z = \slap \psi \ . 
\]
Consider now the equations on $S^2$ 
\bea
 \slap \varphi \ & = & \ f  \ , \label{f1} \\ 
 \slap \psi \ & = & \ g \ ,  \label{g1}
\eea
for sufficiently smooth functions $f, g$ with vanishing mean on $S^2$. 
By the Hodge theorem there exist smooth solutions to (\ref{f1}), respectively (\ref{g1}) that are unique up to an additive constant. 
In our article we consider the function $\varphi$ of vanishing mean of equation (\ref{f1}). Analogously for $\psi$ and equation  (\ref{g1}), where applicable. \\

In this paper, we consider the situation for 
\be \label{Z1}
\mathcal{Z} = \dlap (\Sigma^- - \Sigma^+)
\ee
as well as 
\be \label{Z2}
\mathcal{Z} = \dlap (\Sigma (u) - \Sigma^+) \ . 
\ee 
The above equations determine $(\Sigma^- - \Sigma^+)$, respectively $(\Sigma (u) - \Sigma^+)$ uniquely. \\

\vspace{2cm}



\begin{thebibliography}{99} 
\bibitem{lydia1} L. Bieri.  
        \begin{itshape} An Extension of the Stability Theorem of the Minkowski Space
in General Relativity. \end{itshape}
        ETH Zurich, Ph.D. thesis.  \textbf{17178}. 
        Zurich. (2007).  
\bibitem{lydia2} L. Bieri.  
        \begin{itshape} Extensions of the Stability Theorem of the Minkowski Space
in General Relativity. Solutions of the Einstein Vacuum Equations. \end{itshape}
        AMS-IP. Studies in Advanced Mathematics. Cambridge. MA. (2009).         
\bibitem{lydia3} L. Bieri.  
        \begin{itshape} New Effects in Gravitational Waves and Memory.  \end{itshape} 
Phys. Rev. D 103. 024043. (2021)        
\bibitem{lydia4} L. Bieri.  
        \begin{itshape}  New Structures in Gravitational Radiation.     \end{itshape}
Advances in Theoretical and Mathematical Physics. \textbf{26}. 3. (2022). 531-594. 
See also arxiv (2020). Latest version https://arxiv.org/pdf/2010.07418.pdf 
\bibitem{lbdg2} L. Bieri, D. Garfinkle. 
\begin{itshape}  An electromagnetic analog of gravitational wave memory. \end{itshape} 
Class. Quantum Grav. 30. 19. (2013) 195009.  
\bibitem{lydia5} L. Bieri.  
        \begin{itshape}  Global Solutions to the Einstein Vacuum Equations with Dynamical Term Homogeneous of Degree -1.  \end{itshape}
Preprint. (2022). 
\bibitem{chenkellerwangwangyau} P.-N. Chen, J. Keller, M.-T. Wang, Y.-K. Wang, AND S.-T. Yau. 
\begin{itshape} Evolution of Angular Momentum and Center of Mass at Null Infinity.  
\end{itshape} 
Commun. Math. Phys. 386, 551Ð588 (2021). 
\bibitem{chrmemory}  D. Christodoulou. 
        \begin{itshape} Nonlinear Nature of Gravitation and Gravitational-Wave
Experiments. \end{itshape}
        Phys.Rev.Letters. \textbf{67}. 
        (1991). no.12. 1486-1489. 
\bibitem{chrIV2000}  D. Christodoulou. 
        \begin{itshape} The Global Initial Value Problem in General Relativity. \end{itshape} 
        Proceedings of the 9th Marcel Grossmann Meeting on General Relativity. Rome. Italy. (2000). 
\bibitem{chrdmay2008}  D. Christodoulou. 
        \begin{itshape} The Formation of Black Holes in General Relativity. \end{itshape}
        EMS publishing house ETH Z\"urich. (2009).        
\bibitem{chrdEuler2007}         
       \begin{itshape} The Formation of Shocks in 3-Dimensional Fluids. \end{itshape}
 EMS Monographs in Mathematics. EMS Publishing House. (2007). 
 \bibitem{chrdEuler2019}   
    \begin{itshape} The Shock Development Problem. \end{itshape}
EMS Monographs in Mathematics. EMS Publishing House. (2019).
\bibitem{sta} D. Christodoulou, S. Klainerman.
        \begin{itshape} The global nonlinear stability of the Minkowski space.
\end{itshape}
        Princeton Math.Series \textbf{41}. 
        Princeton University Press. Princeton. NJ. (1993).     
\bibitem{draystreu1} T. Dray and M. Streubel.
\begin{itshape}  Angular momentum at null infinity. \end{itshape}  
Class. Quantum Grav. 1 (1984), no. 1, 15Ð26. 
\bibitem{flanagannichols1} \`E. \`E. Flanagan and D. A. Nichols.
\begin{itshape}   Conserved charges of the extended Bondi-Metzner-
Sachs algebra. \end{itshape}  
 Phys. Rev. D 95, 044002 (2017). 
 \bibitem{arizzi1} A. Rizzi. 
 \begin{itshape}   Angular momentum in general relativity: a new definition. \end{itshape}  
 Phys. Rev. Lett. 81 (1998), no. 6, 1150Ð1153.
\bibitem{zeldovich} Ya.B. Zel'dovich and A.G. Polnarev, Sov. Astron. {\bf 18}, 17 (1974)
\bibitem{zip} N. Zipser. 
        \begin{itshape} The Global Nonlinear Stability of the Trivial Solution of
the Einstein-Maxwell Equations.  \end{itshape}
        Ph.D. thesis. Harvard Univ. Cambridge MA. (2000).          
\bibitem{zip2} N. Zipser.  
        \begin{itshape} Extensions of the Stability Theorem of the Minkowski Space
in General Relativity. - Solutions of the Einstein-Maxwell Equations.
\end{itshape}
        AMS-IP. Studies in Advanced Mathematics. Cambridge. MA. (2009).          
\end{thebibliography}
\end{document}